\documentclass{PoS}
\usepackage{graphicx, amsmath, amssymb, epsf, bm
}
\title{Unintegrated parton densities at small x}

\ShortTitle{Unintegrated parton densities at small x   }

\author{\speaker{Martin Hentschinski}\thanks{Work in collaboration with F.~Hautmann, H.~Jung, A.~Sabio Vera and C.~ Salas Hernandez}\\
        Instituto de F\'isica Te\'orica UAM/CSIC,   Universidad Aut\'onoma de Madrid,  Cantoblanco,  E-28049 Madrid, Spain \\
        E-mail: \email{martin.hentschinski@uam.es}}

      \abstract{We present recent results on transverse momentum
        dependent parton densities in the limit of small Bjorken $x$.
        A definition of an unintegrated seaquark density is proposed
        which can be used within the CCFM Monte-Carlo event generator
        \textsc{Cascade}. We further discuss aspects of the
        determination of a NLO BFKL unintegrated gluon density.  }

\FullConference{The 2011 Europhysics Conference on High Energy Physics-HEP 2011,\\
		July 21-27, 2011\\
		Grenoble, Rh\^one-Alpes France}

\begin{document}

\section{Introduction}
Transverse momentum dependent parton distribution arise naturally at
small $x$ as a consequence of high energy factorization and
BFKL-evolution \cite{bfkl}. A formulation of high energy factorization
which is in accordance with conventional collinear factorization is
provided by $k_T$-factorization as defined in
\cite{ktfac,Catani:1994sq}: at high center of mass energies, the
resummation of high energy logarithms (BFKL) can be brought into a
form consistent with conventional collinear resummation (DGLAP).  In
the following we present ongoing work dedicated to the study of
transverse momentum dependent parton densities at small $x$. In
sec.~\ref{sec:cascade} we address the definition of an unintegrated
sea-quark for the \textsc{Cascade} Monte Carlo event generator.  In
sec.~\ref{sec:nlo} we discuss the determination of a transverse momentum
dependent gluon density which takes into account full next-to-leading
order BFKL resummation.

\section{A $k_T$-dependent sea-quark density for the 
 \textsc{Cascade}}
\label{sec:cascade}

The Monte Carlo event generator \textsc{Cascade} \cite{cascade} is
based on the CCFM evolution equation \cite{ccfm}. The latter
interpolates between DGLAP and BFKL evolution and thus allows
naturally for a Monte Carlo implementation of $k_T$-factorization.
Based on the principle of color coherence, the CCFM parton shower
describes at first only the emission of gluons, while emissions of
quarks are left aside. This is justified as enhanced regions of phase
space for $x\to 1$ and $x\to 0$ are dominated by gluonic dynamics at
leading logarithmic level. As far as unintegrated parton densities are
concerned, this implies that CCFM evolution describes only the distinct
evolution of unintegrated gluon and valence quarks while transitions
between quarks and gluons are left aside. Although formally
sub-leading, quark emissions can give give numerical sizeable
contributions and it is therefore advisable to include them into
account in the parton shower, too.  Absence of quark emissions affects
furthermore the determination of $k_T$-dependent hard matrix elements:
due to absence of a seaquark density, the gluon-to-quark splitting
needs to be included into matrix element which complicates their
determination, see {\it e.g.}  \cite{DYmichal}.  As a first step
towards arriving at a complete inclusion of quark emission into the
CCFM parton shower, we present here a definition of an off-shell
sea-quark density and a partonic matrix element. For the former we
restrict to the case where the gluon-to-quark splitting occurs as the
last evolution step.   For a detailed discussion we refer to
\cite{HHJ}.

For definiteness we consider in the following the transverse momentum
dependent factorization of the $qg^* \to qZ$ matrix element into a
transverse momentum dependent partonic sub-process $qq^* \to Z$ and a
transverse momentum dependent quark-gluon splitting function. The
latter is given by the e $k_T$-dependent quark-gluon splitting
function of \cite{Catani:1994sq} which reads
\begin{align}
  \label{eq:ktsplitt_def}
  {P}_{qg} \bigg(z, \frac{{\bf k}^2 }{ {\bf q}^2 }\bigg) = 
T_R \left( 
            \frac{ {\bf q}^2 }{\,  {\bf q}^2  + z(1-z){\bf k}^2}
\right)^2 
\left[
        {(1-z)^2 + z^2 } + 4z^2 (1-z)^2 \frac{  {\bf k}^2 }{{\bf q}^2}
\right], 
\end{align}
Here ${\bf k}$ and ${\bf q}$ denote the transverse momentum of the
off-shell gluon and off-shell quark respectively, while $z$ denotes
the fraction of gluon light cone momentum which is carried on by the
$t$-channel quark.  Note that in the on-shell limit ${\bf k}^2 \to 0$
eq.~\eqref{eq:ktsplitt_def} reduces to the conventional DGLAP
splitting function $T_R [z^2 + (1-z)^2]$.  For the determination of
the $qq^* \to Z$ coefficient we follow closely the treatment of the
already existing gluonic case. There off-shell gauge invariance is
ensured through a reformulation of QCD at high center of mass energies
in terms of effective degrees of freedom, reggeized gluons. The latter
coincide in their on-shell limit with conventional collinear QCD
gluons. For the general off-shell case  one uses effective
vertices which contain additional induced terms,  which ensure off-shell
gauge invariance\footnote{For an effective action approach to the
  reggeized gluon formalism see \cite{reggeized}.}. In complete
analogy one can construct a reggeized quark formalism for the
description of the high energy limit of scattering amplitudes with
quark exchange in the $t$-channel \cite{reggequarks}. As (reggeized)
quark exchanges are in comparison to (reggeized) gluon exchanges
suppressed by powers of $s$, they generally do not occur in the high
energy resummation of total cross-sections. They can however be used
as a starting point for the construction of an off-shell factorization
of matrix elements which are limited to quark exchange in the
$t$-channel. This allows  to define the off-shell partonic
cross-section  $qq^* \to Z$ as the analytic continuation of the $qq
\to Z$ cross-section, in complete analogy to the gluonic case.  The
splitting function on the other hand is within this approach at first
obtained as a pure constant, corresponding to the limit $z = 0$ of
eq.~(\ref{eq:ktsplitt_def}). It is however possible to incorporate
finite $z$-corrections into the reggeized quark effective vertices,
which allow to re-obtain  eq.~(\ref{eq:ktsplitt_def}) for
the gluon-to-quark splitting. Combination of eq.~\eqref{eq:ktsplitt_def} with the unintegrated gluon density then yields the  unintegrated quark density 
\begin{align}
  \label{eq:seaqark}
  \mathcal{Q}^{\text{sea}} (x, {\bf q}^2, \mu^2) &=   \frac{1}{{\bf q}^2}
   \int\limits_x^1  \frac{d z}{z}
\int \limits_0^{{\bf k}_{\text{max}}^2}
 d {\bf k}^2 
  \frac{\alpha_s(\mu^2)}{2 \pi} {P}_{qg} \left(z, \frac{ {\bf k}^2}  {{\bf q}^2}\right) 
\mathcal{G}\left(\frac{x}{z}, {\bf k}^2, \bar{\mu}^2   \right).
\end{align}
with ${\bf k}^2_{\text{max}} = {  \mu^2}/z - { {\bf q}^2 }/{(z (1-z))} $. For  further details and numerical results we refer to \cite{HHJ}.

\section{NLO BFKL gluon density}
\label{sec:nlo}

We determine a transverse momentum dependent gluon density which
follows NLO BFKL evolution.  It is obtained as the convolution
\begin{align}
  \mathcal{G}(x, {\bm q}^2) & = \int \frac{d^2 {\bm k}}{\pi}  f_{\text{BFKL}}(x, {\bm q}, {\bm k}) \Phi_P({\bm k}^2).
\end{align}
with NLO  BFKL Green's function $f_{\text{BFKL}}$ and a model for the 
proton impact factor
\begin{align}
   \Phi_P({\bm k}^2, Q_0^2,\delta, A) & = 
A \frac{1}{{\bm k}^2}\left(\frac{{\bm k}^2}{Q_0^2} \right)^\delta e^{- \frac{{\bm k}^2}{Q_0^2}},
\end{align}
with free parameters $\{ Q_0, A, \delta \}$ to be obtained from a fit
to HERA data. The $F_2$ structure function requires to convolute the
gluon density with the photon impact factor, for which we use the LO
kinematical improved impact factor of \cite{Bialas:2001ks}. Given the
recent progress in the determination of NLO corrections to effective
vertices of the gluonic effective action \cite{Hentschinski:2011tz},
it seems however possible to obtain NLO impact factors for future
studies of DIS and and LHC cross-sections.  The NLO BFKL Green's
function is known to possess a numerical instability due to the
presence of (anti-) collinear logarithms. For the present study the
latter are resummed using an extension of the `all-pole'-approximation
of \cite{Vera:2005jt}.  It has the advantage that it allows for an
exponentiation of the BFKL kernel, in agreement with
Regge-theory. Furthermore a transformation from transverse moment to
transverse momentum space is possible in that case. This is of
relevance as it allows for an implementation of this Green's function
into a NLO BFKL Monte Carlo event generator \cite{Chachamis:2009ks}
which is currently investigated. For a detailed discussion and the
results of the fit we refer to \cite{HSS}

\subsubsection*{Acknowledgments}

I wish to thank my collaborators F.~Hautmann, H~Jung, A.~Sabio Vera
and C.~Salas Hernandez for intense and numerous discussions.
Financial support from the German Academic Exchange Service (DAAD),
the MICINN under grant FPA2010-17747 and the Research Executive Agency
(REA) of the European Union under the Grant Agreement number
PITN-GA-2010-264564 (LHCPhenoNet) is gratefully acknowledged.

\end{document}